\documentclass[preprint,aps,prc,floatfix,showpacs,showkeys]{revtex4}
\usepackage{graphicx,amsmath,amssymb,dcolumn}
\usepackage{bm}

\newcommand{\be}{$\beta$}

\newcommand{\su}[1]{$^{#1}$}
\newcommand{\msu}[1]{^{#1}}

\newcommand{\msub}[2]{{#1}_{#2}}

\newcommand{\et}{\emph{et al.,}~}

\preprint{}

\begin{document}

\title{Doubly-magic nature of $^{56}$Ni: measurement of the ground state nuclear magnetic dipole moment of $^{55}$Ni}

\author{J.S.~Berryman,$^{1,2}$
K.~Minamisono,$^{2}$
W.F.~Rogers,$^{3}$
B.A.~Brown,$^{2,4}$\\
H.L.~Crawford,$^{1,2}$
G.F.~Grinyer,$^{2}$
P.F.~Mantica,$^{1,2}$
J.B.~Stoker,$^{1,2}$ and
I.S.~Towner$^{5}$
}

\affiliation{$^{1}$
Department of Chemistry, Michigan State University,
East Lansing, Michigan 48824, USA}
\affiliation{$^{2}$ 
National Superconducting Cyclotron
Laboratory, Michigan State University,
East Lansing, Michigan 48824, USA} 
\affiliation{$^{3}$
Department of Physics, Westmont College,
Santa Barbara, CA 93108, USA}
\affiliation{$^{4}$
Department of Physics and Astronomy, Michigan State University, 
East Lansing, Michigan 48824}
\affiliation{$^{5}$
Physics Department, Queen's University,
Kingston, Ontario K7L 3N6, Canada}

\date{\today}
\begin{abstract}

The nuclear magnetic moment of the ground state of $^{55}$Ni ($I^{\pi}=3/2^{-}, T_{1/2}=204$ ms) has been deduced to be $|\mu(^{55}$Ni)$|=(0.976 \pm 0.026)$ $\mu_N$ using the $\beta$-NMR technique. Results of a shell model calculation in the full \textit{fp} shell model space with the GXPF1 interaction reproduce the experimental value.  Together with the known magnetic moment of the mirror partner $^{55}$Co, the isoscalar spin expectation value was extracted as $\langle \sum \sigma_z \rangle=0.91 \pm 0.07$.  The $\langle \sum \sigma_z \rangle$ shows a similar trend as that established in the \textit{sd} shell.  The present theoretical interpretations of both $\mu(^{55}$Ni) and $\langle \sum \sigma_z \rangle$ for the $T=1/2$, $A=55$ mirror partners support the softness of the $^{56}$Ni core.

\end{abstract}


\maketitle

\section{Introduction}

The nuclear magnetic dipole moment sensitively reflects which single-particle orbits contribute to the nuclear wave function, yielding key information on nuclear shell structure, especially shell evolution and shell closures (magicity).  The magnetic moments of nuclei one nucleon removed from doubly-closed shells are of particular importance, since the properties of the nucleus should be determined by the orbit occupied by the unpaired nucleon.  Deviations in these properties from theory may indicate the presence of higher-order configuration mixing, meson exchange currents (MEC), isobar excitation, and$/$or even a breakdown of the magicity.

The character of stable nuclei with magic numbers of both protons and neutrons, such as \su{16}O and \su{40}Ca, has been well established.  The radioactive doubly magic nuclei, however, have revealed interesting surprises.  An extreme example is that of \su{28}O, which was expected to be bound based on its doubly-magic character (proton and neutron numbers $Z=8$ and $N=20$, respectively), but has been shown to be unbound \cite{sakurai}.  The study of $\beta$ unstable \su{56}Ni, residing three neutrons away from stability, may provide insight into changes in the structure of doubly-magic nuclei as one moves further from stability.  All eight magnetic moments of the doubly-closed shell \su{16}O and \su{40}Ca $\pm$ 1 nucleon nuclei are experimentally known \cite{tanigaki, bald, alder, tminamisono1, tminamisono2, kusch, brun, tminamisono3} and agree well with the values obtained assuming an inert core $\pm$ 1 nucleon (single-particle value).  The agreement reflects the ``goodness'' of the \su{16}O and \su{40}Ca cores.  The nucleus \su{56}Ni is the first self-conjugate nucleus with magic neutron and proton numbers ($N=Z=28$) that is radioactive.  The three known magnetic moments around \su{56}Ni \cite{Min,Cal,Oht} do not agree with single-particle values.  The discrepancy indicates the necessity of corrections to the simple picture of a \su{56}Ni closed shell, where the \su{56}Ni core is described by the lowest order configuration of nucleons plus a sizable mixture of other configurations, in other words, the \su{56}Ni core is soft.

The nuclei one nucleon away from \su{56}Ni are: \su{55}Ni (neutron hole in $1f_{7/2}$), \su{55}Co (proton hole in $1f_{7/2}$), \su{57}Cu (proton particle in $2p_{3/2}$), and \su{57}Ni (neutron particle in $2p_{3/2}$).  The measured magnetic moments of \su{55}Co \cite{Cal} and \su{57}Ni \cite{Oht}, isospin projection $T_z=+1/2$ nuclei, are well reproduced by the large scale shell model calculation in the full \textit{fp} shell using the GXPF1 interaction \cite{Hon}.  The experimental results support \su{56}Ni as being a soft core as the probability of the $N=Z=28$ lowest order closed shell configuration is 60\%.  The magnetic moment of the $T_z=-1/2$ nucleus \su{57}Cu was measured to be $|\mu$(\su{57}Cu)$|=$ (2.00 $\pm$ 0.05) $\msub{\mu}{N}$ \cite{Min}.  The shell-model calculation gives $\mu$(\su{57}Cu)= +2.45 $\msub{\mu}{N}$, and disagrees with the experimental value. The large discrepancy between experiment and theory for $\mu$(\su{57}Cu) suggests an even softer core, or a major shell breaking at \su{56}Ni.  In the present study, $\mu$ of the $T_z=-1/2$ nucleus \su{55}Ni was measured for the first time using the \be-ray detecting nuclear magnetic resonance (\be-NMR) technique to probe the structure of \su{56}Ni using the one neutron hole in the $1\msub{f}{7/2}$ shell.  

The softness of the \su{56}Ni core also appears in the contradicted behavior between the first excited 2\su{+} state and the reduced transition matrix element, $B(E2;0_1^+ \to 2_1^+)$, within the Ni isotopic chain.  The energy of the $2_1^+$ state in \su{56}Ni, $E(2_1^+)=2701$ keV, is significantly higher than those of its neighboring even-even nuclei, suggesting a good \su{56}Ni core.  However, the adopted value of $B(E2;0_1^+ \to 2_1^+)=(600 \pm 120) e^2$ fm\su{4} \cite{raman} of \su{56}Ni does not show significant variation from those of nearest neighbor isotopes.  A reduced $B(E2;0_1^+ \to 2_1^+)$ at \su{56}Ni would be expected for a good core.  The disparate nature of the $E(2_1^+)$ and $B(E2;0_1^+ \to 2_1^+)$ in \su{56}Ni was explained by a large scale shell model calculation with the quantum Monte Carlo diagonalization method in the full \textit{fp} shell \cite{otsuka}.  The calculation reproduced the experimentally-observed $E(2_1^+)$ and $B(E2;0_1^+ \to 2_1^+)$ using the FPD6 interaction, wherein the probability of the $N=Z=28$ lowest order closed shell component in the wavefunction of the \su{56}Ni ground state was only 49\%, compared to an 86\% of the closed shell component in the wavefunction of the \su{48}Ca ground state.  

\section{Experimental Procedure}

The $\beta$-NMR measurement on \su{55}Ni was performed at National Superconducting Cyclotron Laboratory (NSCL) at Michigan State University.  The \su{55}Ni ions were produced from a primary beam of \su{58}Ni accelerated up to 160 MeV$/$nucleon by the coupled cyclotrons and impinged on a 610 mg$/$cm\su{2} \su{9}Be target.  The primary beam was set at an angle of $+2\msu{\circ}$ relative to the normal beam axis at the production target to break the symmetry of the fragmentation reaction and produce a nuclear spin-polarized beam of \su{55}Ni \cite{asahi}.   The A1900 \cite{Mor} was used for the initial separation of the \su{55}Ni from other reaction products using the full angular acceptance ($\pm2.5\msu{\circ}$).  An achromatic aluminum wedge (405 mg$/$cm\su{2}) was placed at the second dispersive image of the A1900 for a separation of \su{55}Ni based on relative energy loss.  The momentum acceptance was 1\% and the measurement was performed at +0.14\% momentum relative to the peak of the momentum distribution.  A typical counting rate of \su{55}Ni ions at the \be-NMR apparatus was about 500 particles/s/pnA, with the primary beam set at $2\msu{\circ}$ and intensity 5 pnA.  The major contamination in the secondary beam following the A1900 was \su{54}Co ($I^{\pi}=0^+$, $T_{1/2}=193.3$ ms), which had a similar magnetic rigidity as \su{55}Ni. Significant \su{54}Co contamination can negatively impact the \su{55}Ni $\beta$-NMR measurement due to its similar half-life and \be-endpoint energy to those of \su{55}Ni.  Therefore, the Radio-frequency Fragment Separator (RFFS) \cite{Gor} was used to remove \su{54}Co from the secondary beam based on the time of flight difference, and a beam purity $>99$\% in \su{55}Ni was realized in the experiment.

The polarized \su{55}Ni ions were implanted into a NaCl single crystal with a cubic lattice structure at the center of the \be-NMR apparatus \cite{Man}.  The apparatus consisted of a dipole magnet with its poles perpendicular to the beam direction and a 10 cm pole gap.  The magnet provided the required Zeeman splitting of the nuclear-magnetic levels of the spin-polarized nuclei.  Two $\beta$ telescopes, each consisting of a thin (4.4 cm $\times$ 4.4 cm $\times$ 0.3 cm) and a thick (5.1 cm $\times$ 5.1 cm $\times$ 2.5 cm) plastic scintillator, were placed at 0$^{\circ}$ and 180$^{\circ}$, relative to the direction of polarization, between the poles of the magnet. Two identical radiofrequency (\textit{rf}) coils in a Helmholtz-like geometry were placed within the magnet and the $\beta$ telescopes, and made up an LCR resonance circuit \cite{minamisono_nqr} where $L$ is the inductance of the \textit{rf} coil, $C$ is the capacitance, and $R$ is the resistance.  The magnetic field created by the \textit{rf} coils was perpendicular to both the direction of the beam and the static magnetic field.  The 20-mm diameter and 2-mm thick NaCl crystal was mounted on an insulated holder between the pair of \textit{rf} coils and the face of the crystal was tilted at $45\msu{\circ}$ relative to the direction of the beam and the poles of the magnet to reduce \be-ray scattering in the crystal.

\su{55}Ni decays to the ground state of \su{55}Co emitting \be\su{+} particles with a half-life of 204 ms.  The branching ratio to the ground state ($I\msu{\pi}=7/2\msu{-}$) is 100\% and the maximum $\beta$ energy is 7.67 MeV.  The data acquisition was triggered each time a coincidence event was registered between the thin and thick $\beta$ detectors of either telescope.  Because of the asymmetric $\beta$-ray angular distribution from the polarized nuclei, $W(\theta)\sim1+A_{\beta}P$cos$\theta$, the counting rates between the $0^{\circ}$ and $180^{\circ}$ counters were asymmetric depending on the \be-decay asymmetry parameter $A_{\beta}$, $P$, and the angle $\theta$ between the momentum direction of the $\beta$ and the polarization axis.

An independent asymmetry measurement that deduces the magnitude of spin polarization as well as direction was performed to compare to the magnitude and direction of the NMR effect observed in the $\beta$-NMR measurement. A technique has been developed at NSCL to measure polarization using a pulsed external magnetic field and does not require advanced knowledge of the nuclide's magnetic moment \cite{anthony}.  When the external magnetic field, $H_0$, is on, the spin polarization is maintained in the crystal and the $\beta$ angular distribution is asymmetric.  When $H_0$ is off, spin polarization is not maintained.  The $H_0$ was set at 1000 G and the pulse duration was 60 s on and 60 s off in a repetitive cycle.  The asymmetry change, $A_{\beta}P$, extracted from the ratio of $W(\theta)$ between $H_0$ off and on, 
\begin{equation}
A_{\beta}P=\frac{R-1}{R+1},
\end{equation}
where
\begin{equation} \label{eq:doubleratio}
R=\frac{[\msub{W(0^{\circ})/W(180^{\circ})]}{\textrm{off}}}{[\msub{W(0^{\circ})/W(180^{\circ})]}{\textrm{on}}}
\end{equation}
was measured at a primary beam angle of 2$^{\circ}$.  However, $R$ also reflects any instrumental asymmetries, for example, the effect of $H_0$ on$/$off on the photomultiplier tubes used to detect the $\beta$ particles.  A normalization for $R$ was provided by separate measurements of $R$ with the secondary beam at 0$^{\circ}$ along the incident beam direction, where no polarization was produced, to correct for this spurious asymmetry. 

The NMR measurement was performed with $\msub{H}{0}=(0.4491 \pm 0.0005)$ T, measured by a proton-NMR magnetometer.  An \textit{rf} on and off technique with continuous \su{55}Ni implantation was employed.  A frequency-modulated \textit{rf} was applied to the \su{55}Ni in NaCl for a duration of 30 s on and 30 s off in a repetitive cycle.  Typical frequency modulation (FM), \textit{rf} time to sweep the FM, and amplitude were $\pm$25 kHz, 20 ms, and 0.7 mT, respectively.  $A_{\beta}P$ was extracted from the ratio in Eq.\ (\ref{eq:doubleratio}) for \textit{rf} off and \textit{rf} on, and was measured as a function of the applied frequency, $\nu$.  The $g$ factor was extracted from the resonance frequency (Larmor frequency) $\nu=\nu_L$ with:
\begin{equation}
h\nu_L=g \mu_N H_0.
\end{equation}
All measurements were performed at room temperature. 

\section{Results}

\begin{figure}
\includegraphics[width=8cm]{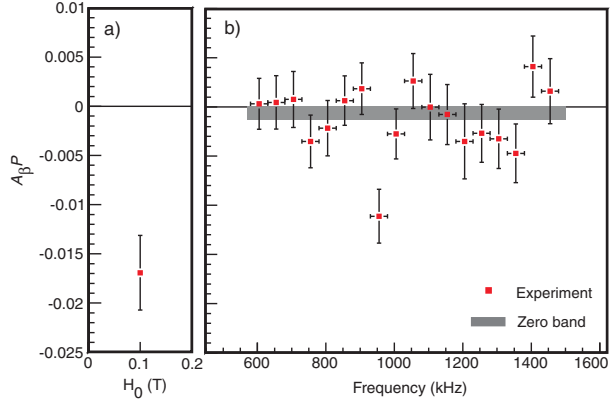}
\caption{(Color online) a) $A_{\beta}P$ measured by the $H_0$ on$/$off technique.  b) NMR spectrum of \su{55}Ni in NaCl, where $A_{\beta}P$ was determined as a function of applied \textit{rf}.  Data were taken at $H_0=(0.4491 \pm 0.0005)$ T and with FM=$\pm25$ kHz shown as a horizontal bar at each point.  The squares are the experimental values and the band is the base line obtained from a weighted average of all the data except the resonance point at 955 kHz.}
\label{fig:nmr_final}
\end{figure}

The result of the spin polarization measurement is shown in Fig.\ \ref{fig:nmr_final}a), where $A_{\beta}P$ is plotted at the value of the external magnetic field used for the measurement.  $A_\beta$ may have one of two values, $A_{\beta}=+0.885$ or -0.747 \cite{morita} depending on the sign of the mixing ratio $\rho=\frac{C_A\langle\sigma\rangle}{C_V\langle1\rangle}$, where $C_V$ and $C_A$ are the vector and axial-vector coupling constants, $\langle1\rangle$ is the Fermi matrix element, and $\langle\sigma\rangle$ is the Gamow-Teller matrix element.  The two values for $A_{\beta}$ are similar in magnitude and the absolute value for spin polarization was extracted as $|P|\approx 2$\%.

The resulting NMR spectrum is shown in Fig.\ \ref{fig:nmr_final}b), where $A_{\beta}P$ is plotted as a function of applied \textit{rf}.  The resonance was found at frequency $\nu_L=955$ kHz with FM$=\pm 25$ kHz.  The magnitude and sign of $A_{\beta}P$ at the resonance frequency are consistent with those obtained in the spin polarization measurement shown in Fig.\ \ref{fig:nmr_final}a).  The confidence interval for the mean of the baseline was determined, and compared to the statistical error in $A_{\beta}P$ at 955 kHz.  At the 95\% confidence level, the 955 kHz point lies 3$\sigma$ from the baseline.  The corresponding $g$ factor was deduced as $|g|=0.279 \pm 0.007$.  The magnetic moment can be further extracted as $\mu=gI$, with $I=7/2$ for the \su{55}Ni ground state \cite{aysto}.  The final result is 
\begin{displaymath}
|\mu(\msu{55}\textrm{Ni})|=(0.976 \pm 0.026) \msub{\mu}{N}.
\end{displaymath}
The uncertainty on $\mu$ was evaluated from the FM. The $\mu$ was not corrected for the chemical shift, which is not known, but assumed to be small compared to the error on the present result.  The sign of $g$ and thus $\mu$ cannot be determined directly from the measurement.  However, it was assumed negative based on theoretical considerations for a neutron hole in the $1\msub{f}{7/2}$ shell. 

\section{Discussion}

\subsection{Magnetic moment with single-particle wavefunction}
The new $\mu$(\su{55}Ni) was first compared to the results of a calculation that used a simple form of the wavefunction, where \su{56}Ni was assumed to be an inert closed core, with a description of the magnetic moment operator $\vec{\mu}_{\textrm{eff}}=g_{l,\textrm{eff}} \langle l \rangle +g_{s,\textrm{eff}}\langle s \rangle+g_{p,\textrm{eff}}\langle[Y_2,s]\rangle$, where $g_{x,\textrm{eff}}=g_x+\delta g_x$, with $x=l,s$, or $p$ \cite{castel,towner}, and $g_p$ denotes a tensor term.  Here $g_x$ is the free nucleon $g$ factor $g_\textrm{free}$ ($g_s=5.586$, $g_l=1$ for proton and $g_s=-3.826$, $g_l=0$ for neutron) and $\delta g_x$ the correction to it.  $s$ and $l$ represent spin and orbital angular momentum, respectively.  The perturbation calculation applied corrections for core polarization (CP) and meson exchange currents (MEC).  CP is a correction to the single-particle wavefunction that occurs when there is an excitation in the closed-shell core made from a particle in orbital ($l-s$) coupling to a hole in orbital ($l+s$).  MEC corrections applied to the magnetic moment operator account for nucleons interacting via the exchange of charged mesons.  Details of the calculation and individual corrections can be found in Refs.\ \cite{castel,towner,golovko}.  Starting from the single-particle values for \su{55}Ni [$\mu$(\su{55}Ni)=-1.913 $\mu_N$], whose magnitude is larger than experiment, the CP corrections overcorrect experimental values [$\mu$(\su{55}Ni)=-0.169 $\mu_N$ with CP only], but the MEC corrections restore the calculated value toward experiment [$\mu$(\su{55}Ni)=-1.235 $\mu_N$ with CP+MEC].  Including additional relativistic and isobar corrections \cite{castel,towner,golovko}, the simple theoretical model reproduces the experimental value well, as shown in Table \ref{tab:values}, labeled as $g_\textrm{eff}^\textrm{perturbation}$, together with the results for the mirror partner \su{55}Co.
\begin{table*}
\caption{Magnetic moments of \su{55}Ni,\su{55}Co and the isoscalar spin expectation values of the mass $A=55$ system.}
\label{tab:values}
\begin{ruledtabular}
\begin{tabular}{lccc}
    &    \textbf{$\mu$(\su{55}Ni) $\msub{\mu}{N}$}   &   \textbf{$\mu$(\su{55}Co) $\msub{\mu}{N}$}   &   \textbf{$\left\langle\sum\msub{\sigma}{z}\right\rangle$}  \\
\hline
Experiment & $-0.976 \pm 0.026$  & $4.822 \pm 0.003$ \cite{Cal} & $0.91 \pm 0.07$ \\
\hline
Single-particle value   & -1.913 & 5.792 & 1.00 \\
$g_\textrm{eff}^\textrm{perturbation}$ (see Refs.\ \cite{castel,towner,golovko} for details) & -1.072 & 4.803 & 0.61 \\
\hline
full \textit{fp} $g_{\textrm{free}}$ & -0.809 & 4.629 & 0.84 \\
full \textit{fp} $g_{\textrm{eff}}^\textrm{moments}$  & -0.999 & 4.744 & 0.65 \\
full \textit{fp} $g_\textrm{eff}^{sd \textrm{ fit}}$ & -1.071 & 4.926  & 0.94 \\
full \textit{fp} $g_\textrm{eff}^{sd \textrm{ fit}}$ without isoscalar $\delta g_l^{sd \textrm{ fit}}$ term & -1.129 & 4.868 & 0.63 \\
\end{tabular}
\end{ruledtabular}
\end{table*}

\subsection{Magnetic moment with shell-model wavefunction}
Another theoretical approach was taken using a complex wavefunction in a shell model calculation to gain more insight on the details of the \su{56}Ni core.  The shell model calculation was performed in the full \textit{fp} shell with the effective interaction GXPF1 \cite{Hon}, where \su{40}Ca was assumed to be an inert closed core.  Here, the \su{56}Ni core is soft as the probability of the lowest order closed-shell $\pi(1f_{7/2})^8 \nu(1f_{7/2})^8$ configuration in the ground-state wavefunction is $\sim$60\%.  The magnetic moment can be calculated from $g_\textrm{free}$ with a form of the magnetic moment operator $\vec{\mu}=g_s \langle s \rangle +g_l \langle l \rangle$.  In general, good agreement is realized by this treatment for $N \sim Z$ nuclei over the range $A=47-72$.  The shell model calculation gives $\mu$(\su{55}Ni)=-0.809 $\msub{\mu}{N}$ with $g_\textrm{free}$, which is in fair agreement with the present result as compared with other $\mu$ calculations in Ref.\ \cite{Hon} and supports the softness of the \su{56}Ni core.  Similar results were obtained for the probability of the $\pi(1f_{7/2})^8 \nu(1f_{7/2})^8$ closed shell component in the wavefunction from a separate shell model calculation \cite{otsuka} that explained the discrepancy between the systematics of $E(2_1^+)$ and that of $B(E2;0_1^+ \to 2_1^+)$. 

Effective nucleon $g$ factors, $g_\textrm{eff}^\textrm{moments}$, may be employed in the shell model calculation for better agreement.  The $g_\textrm{eff}^\textrm{moments}$ were derived empirically by the least-square fit of the magnetic moment operator to experimental $\mu$(\su{57-65,67}Ni) and $\mu$(\su{62-68,70}Zn) \cite{Hon}.   The values $g_\textrm{eff}^s=0.9 g_\textrm{free}^s$, $g_\textrm{eff}^l=1.1$ and -0.1 for protons and neutrons, respectively, were obtained.  The resulting magnetic moment, $\mu$(\su{55}Ni)=-0.999 $\msub{\mu}{N}$, gives good agreement with the experimental value.  The results of the theoretical calculations are summarized in Table \ref{tab:values}.  It is noted that all of the theoretical calculations give good agreement with the experimental value, and within the accuracy of nuclear structure models, there is not a significant difference between the result of the calculations for $\mu$.

\subsection{Isoscalar spin expectation value}
Examination of only the contribution from nuclear spins to the magnetic moment can also provide insight into shell structure and configuration mixing.  The magnetic moment can be expressed as the sum of the expectation values of isoscalar $\left\langle\sum\msub{\mu}{0}\right\rangle$ and isovector $\left\langle\sum\msub{\mu}{z}\right\rangle$ components, assuming isospin is a good quantum number, as
\begin{eqnarray}
\mu&=&\left\langle\sum\msub{\mu}{0}\right\rangle+\left\langle\sum\msub{\mu}{z}\right\rangle \\
&=&\left\langle\sum\frac{\msub{l}{z}+(\msub{\mu}{p}+\msub{\mu}{n})\msub{\sigma}{z}}{2}\right\rangle \nonumber\\
& & + \left\langle\sum\frac{\msub{\tau}{z}[\msub{l}{z}+(\msub{\mu}{p}-\msub{\mu}{n})\msub{\sigma}{z}]}{2}\right\rangle,
\end{eqnarray}
where $l$ and $\sigma=2s$ are the orbital and spin angular-momentum operators of the nucleon, respectively, $\tau$ is the isospin operator, $\msub{\mu}{p}=2.793 \mu_N$ and $\msub{\mu}{n}=-1.913 \mu_N$ are the magnetic moments of the free proton and neutron, respectively, and the sum is over all nucleons. The isovector $\left\langle\sum\msub{\mu}{z}\right\rangle$ component depends on the isospin, $\msub{\tau}{z}$, and changes its sign for $T_z=\pm T$.  The isoscalar spin expectation value $\left\langle\sum\msub{\sigma}{z}\right\rangle$ can be extracted from the sum of mirror pair magnetic moments as
\begin{equation} \label{eq:spinexp}
\left\langle\sum\msub{\sigma}{z}\right\rangle = \frac{\mu(T_z=+T)+\mu(T_z=-T)-I}{\msub{\mu}{p}+\msub{\mu}{n}-1/2},
\end{equation}
where the total spin is $I = \left\langle\sum\msub{l}{z}\right\rangle+\left\langle\sum\msub{\sigma}{z}\right\rangle/2$. $\left\langle\sum\msub{\sigma}{z}\right\rangle$ amplifies small differences in theoretical $\mu(T_z=+T)$ and $\mu(T_z=-T)$ and thus is more sensitive to small changes in the magnetic moments of the mirror pair.

Sugimoto \cite{Sug} and later Hanna and Hugg \cite{Han} analyzed data on magnetic moments for mirror nuclei, and found regularities in the spin expectation values for nuclei in the \textit{sd} shell.  All of the ground state magnetic moments of $T=1/2$ mirror nuclei have been measured in the \textit{sd} shell and a systematic trend has been established.  The values of  $\left\langle\sum\msub{\sigma}{z}\right\rangle$ are close to the single-particle value at the beginning of a major shell, and decrease approximately linearly with mass number, reflecting core polarization effects.  In the \textit{fp} shell, however, only three mirror pairs have been measured, masses $A$=41, 43, and 57, and no systematic behavior has been established.

The existing data for $\mu$(\su{55}Co) was combined with the present result to extract $\left\langle\sum\msub{\sigma}{z}\right\rangle$ for the mirror pair at $A=55$.  Using Eq.\ (\ref{eq:spinexp}), 
\begin{displaymath}
\left\langle\sum\msub{\sigma}{z}\right\rangle=0.91 \pm 0.07
\end{displaymath} 
was obtained.  A peculiar feature is noted in Table \ref{tab:values} between experimental and theoretical $\mu$ and $\left\langle\sum\msub{\sigma}{z}\right\rangle$ for $A=55$. Reasonable agreement is achieved among all calculations for $\mu$, but there is variation in the result for theoretical $\left\langle\sum\msub{\sigma}{z}\right\rangle$.  Such feature was already noted in the \textit{sd} shell, and can be explained by examining the isovector and isoscalar components of the M1 operator separately \cite{brown1,brown2}.  The magnetic moment is dominated by the isovector term due to the opposite signs and nearly equal magnitude of the neutron and proton magnetic moments, whereas $\left\langle\sum\msub{\sigma}{z}\right\rangle$ is an isoscalar quantity.  Therefore, small differences in $\mu$ are amplified in $\left\langle\sum\msub{\sigma}{z}\right\rangle$.

To see if a similar approach would realize success in the \textit{fp} shell, the effective $g$ factors for the $A=28$ system obtained from a fit to isoscalar magnetic moments, isovector moments, and M1 decay matrix elements \cite{brown2}, $g_\textrm{eff}^{sd \textrm{ fit}}$, were applied to matrix elements for $A=55$ calculated in Ref. \cite{Hon} with the GXPF1 interaction.  This approach assumes the hole configuration in the 1$d_{5/2}$ shell is analogous to that of 1$f_{7/2}$.  Effective $g$ factors for $A=28$ were obtained as $g_s^{sd \textrm{ fit}} = 4.76$, -3.25, $g_l^{sd \textrm{ fit}}=1.127$, -0.089 and $(g_p')^{sd \textrm{ fit}}=0.41$, -0.35 for protons and neutrons, respectively ($g_p'=g_p/ \sqrt{8\pi}$).   The calculated $\left\langle\sum\msub{\sigma}{z}\right\rangle$ = 0.935 with $g_\textrm{eff}^{sd \textrm{ fit}}$ shows the best agreement with the present result as summarized in Table \ref{tab:values}.  

The \su{56}Ni core could be considered as a good core since $\left\langle\sum\msub{\sigma}{z}\right\rangle$ for $A=55$ is very close to the single-particle value.  However, if the \su{56}Ni core is soft as shown from the satisfactory $\mu$ results from the shell model calculation with the GXPF1 interaction, then configuration mixing should account for the $\sim$40\% of the ground state wavefunction not attributed to $\pi(1f_{7/2})^8 \nu(1f_{7/2})^8$.  This configuration mixing should appear as a deviation in $\left\langle\sum\msub{\sigma}{z}\right\rangle$ from the single-particle value, which is not observed.  It can be shown from the $\left\langle\sum\msub{\sigma}{z}\right\rangle$=0.628 calculated without isoscalar correction to the $g_l^{sd \textrm{ fit}}$, $\delta_l^{IS}$, that a contribution from the large orbital angular momentum ($f$ orbit) to the $g_l^{sd \textrm{ fit}}$ enhances the $\left\langle\sum\msub{\sigma}{z}\right\rangle$.  The contribution to $\left\langle\sum\msub{\sigma}{z}\right\rangle$ from the large orbital angular momentum correction cancels the effect from configuration mixing, supporting the softness of the \su{56}Ni core and emphasizing the sensitivity of $\left\langle\sum\msub{\sigma}{z}\right\rangle$ to nuclear structure.  Similar enhancement of $\left\langle\sum\msub{\sigma}{z}\right\rangle$ due to $\delta_l^{IS}$ was found in Fig.\ 5 of Ref.\ \cite{brown2} for $A=39$.  The enhancement may be attributed to a large MEC contribution to $\delta_l^{IS}$.   Calculations by Arima \et \cite{arima} that included MEC corrections were found to agree with the empirical value of $\delta_l^{IS}$.  However, it is noted that the MEC depends sensitively on the choice of the meson-nucleon coupling constants (see Ref.\ \cite{brown2,towner}) and that calculations by Towner \cite{towner} do not show such enhancement, attributed to the MEC being offset by the relativistic effect.  The contribution to $\left\langle\sum\msub{\sigma}{z}\right\rangle$ from the tensor term $g_p^{sd \textrm{ fit}}$ is small as $\left\langle\sum\msub{\sigma}{z}\right\rangle$=0.94 (0.87) is calculated with (without) the tensor term.  The good agreement between the present result and the $\left\langle\sum\msub{\sigma}{z}\right\rangle$ calculated with $g_{\textrm{eff}}^{sd \textrm{ fit}}$ in the \textit{sd} shell implies that a universal operator can be applied to both the \textit{sd} and \textit{fp} shells.  However, for more detailed discussion, effective M1 operators of the \textit{fp} shell nuclei have to be determined from the mirror moments in the \textit{fp} shell, for which more experimental data are required.  

\subsection{Buck-Perez analysis}
Finally, our result can also be compared to the predictions made by Buck \et \cite{Buc1,Buc2,Per} based on the systematic linear relationship between ground state $g$ factors and the $\beta$-decay transition strengths of mirror nuclei.  The predicted values are $\mu$(\su{55}Ni)=(-0.945 $\pm$ 0.039) $\mu_N$ based on the linear trend of experimental $g$ factors and $\mu$(\su{55}Ni)=(-0.872 $\pm$ 0.081) $\mu_N$ based on the dependence of $ft$ values.  Both predictions are in agreement with the experimental value [$\mu$(\su{55}Ni)=(-0.976 $\pm$ 0.026) $\mu_N$].  The Buck-Perez systematic relation is a valid prediction for \textit{fp} shell nuclei with unknown magnetic moments, and an important tool for future measurements.   

\section{Conclusion}
The magnetic moment of the $T=1/2$ nucleus \su{55}Ni was deduced for the first time as $|\mu$(\su{55}Ni)$|=(0.976 \pm 0.026)$ $\msub{\mu}{N}$.  The experimental result agrees with shell model calculations with the GXPF1 interaction in the full \textit{fp} shell.  The spin expectation value was extracted together with the known $\mu$(\su{55}Co) as $\left\langle\sum\msub{\sigma}{z}\right\rangle=0.91 \pm 0.07$.  The effective $g$ factors determined by isoscalar magnetic moments, isovector moments, and M1 decay matrix elements in the \textit{sd} shell combined with $A=55$ matrix elements are able to explain the present $\left\langle\sum\msub{\sigma}{z}\right\rangle$.  The agreement implies that a universal operator can be applied to both the \textit{sd} and \textit{fp} shells.  The present $\mu$ and $\left\langle\sum\msub{\sigma}{z}\right\rangle$ support the softness of the \su{56}Ni core.  Continued studies of magnetic moments of nuclei immediately outside of presumed doubly-magic cores are important in the ongoing investigation of the resilience of the magic numbers further from stability.

\section{Acknowledgements}
The work was supported in part by the National Science Foundation (NSF) grants PHY-06-06007 and PHY-07-58099.  The authors thank the NSCL operations staff for providing the beams for this experiment.  JSB acknowledges support from the NSF Graduate Research Fellowship program.


\begin{thebibliography}{99}

\bibitem{sakurai}
H. Sakurai \emph{et al.,}~ Phys. Lett. B \textbf{448}, 180 (1999).

\bibitem{tanigaki}
M. Tanigaki \emph{et al.,}~ Hyp. Interact. \textbf{78}, 105 (1993).

\bibitem{bald}
J.D. Baldeschwieler, J. Chem. Phys. \textbf{3}, 152 (1962).

\bibitem{alder}
F. Alder and F.C. Yu, Phys. Rev. \textbf{81}, 1067 (1951).

\bibitem{tminamisono1}
T. Minamisono \emph{et al.,}~ Hyp. Interact. \textbf{78}, 111 (1993).

\bibitem{tminamisono2}
T. Minamisono, J.W. Hugg, D.G. Mavis, T.K. Saylor, H.F. Glavish, and S.S. Hanna, Phys. Lett. B \textbf{61}, 155 (1976).

\bibitem{kusch}
P. Kusch, S. Millman, and I.I. Rabi, Phys. Rev. \textbf{55}, 1176 (1939).

\bibitem{brun}
E. Brun, J.J. Kraushaar, W.L. Pierce, and Wm.J. Veigele, Phys. Rev. Lett. \textbf{9}, 166 (1962).

\bibitem{tminamisono3}
T. Minamisono, Y. Nojiri, K. Matsuta, K. Takeyama, A. Kitagawa, T. Ohtsubo, A. Ozawa, and M. Izumi, Nucl. Phys. A \textbf{516}, 365 (1990).

\bibitem{Min}
K. Minamisono \emph{et al.,}~ Phys. Rev. Lett. \textbf{96}, 102501 (2006).

\bibitem{Cal}
P.T. Callaghan, M. Kaplan, and N.J. Stone, Nucl. Phys. \textbf{A201}, 561 (1973).

\bibitem{Oht}
T. Ohtsubo, D.J. Cho, Y. Yanagihashi, S. Ohya, and S. Muto, Phys. Rev. C \textbf{54}, 554 (1996).

\bibitem{Hon}
M. Honma, T. Otsuka, B.A. Brown, and T. Mizusaki, Phys. Rev. C \textbf{69}, 034335 (2004).

\bibitem{raman}
S. Raman, C.W. Nestor, Jr. and P. Tikkanen, At. Data Nucl. Data Tables \textbf{78}, 1 (2001).

\bibitem{otsuka}
T. Otsuka, M. Honma, and T. Mizusaki,  Phys. Rev. Lett. \textbf{81}, 1588 (1998).

\bibitem{asahi}
K. Asahi \emph{et al.,}~ Phys. Lett. B \textbf{251}, 488 (1990).

\bibitem{Mor}
D.J. Morrissey, B.M. Sherrill, M. Steiner, A. Stolz, and I. Wiedenhoever, NIMB \textbf{204}, 90 (2003).

\bibitem{Gor}
D. Gorelov, V. Andreev, D. Bazin, M. Doleans, T. Grimm, F. Marti, J. Vincent, and X. Wu, Proceedings of 2005 Particle Accelerator Conference, ed. C Horak (IEEE Publishing, Piscataway, New Jersey, (2005) p. 3880.

\bibitem{Man}
P.F. Mantica, R.W. Ibbotson, D.W. Anthony, M. Fauerbach, D.J. Morrissey, C.F. Powell, J. Rikovska, M. Steiner, N.J. Stone, and W.B. Walters, Phys. Rev. C \textbf{55}, 2501 (1997).

\bibitem{minamisono_nqr}
K. Minamisono, R.R. Weerasiri, H.L. Crawford, P.F. Mantica, K. Matsuta, T. Minamisono, J.S. Pinter, and J.B. Stoker, Nucl. Instrum. Methods Phys. Res., Sect. A \textbf{589}, 185 (2008).

\bibitem{anthony}
D.W. Anthony, P.F. Mantica, D.J. Morrissey, and G. Georgiev. \emph{Hyperfine Interact.}, \textbf{127}, 485 (2000).

\bibitem{morita}
M. Morita, \emph{Beta decay and muon capture}, Benjamin, Massachusetts, 1973.

\bibitem{aysto}
J. \"Ayst\"o, J. \"Arje, V. Koponen, P. Taskinen, H. Hyv\"onen, A. Hautoj\"arvi, and K. Vierinen, \emph{Phys. Lett. B}, \textbf{138}, 369 (1984).

\bibitem{castel}
B. Castel and I.S. Towner, \emph{Modern Theories of Nuclear Moments} (Clarendon Press, Oxford, 1990).

\bibitem{towner}
I.S. Towner, Phys. Rep. \textbf{155}, 263 (1987).

\bibitem{golovko}
V.V. Golovko \emph{et al.,}~ Phys. Rev. C \textbf{70}, 014312 (2004).

\bibitem{Sug}
K. Sugimoto,  Phys. Rev. \textbf{182}, 1051 (1969).

\bibitem{Han}
S.S. Hanna and J.W. Hugg,  Hyp. Int. \textbf{21}, 59 (1985).

\bibitem{brown1}
B.A. Brown and B.H. Wildenthal, Phys. Rev. C \textbf{28}, 2397 (1983).

\bibitem{brown2}
B.A. Brown and B.H. Wildenthal, Nucl. Phys. \textbf{A474}, 290 (1987).

\bibitem{arima}
A. Arima, K. Shimizu, W. Bentz, and H. Hyuga, \emph{Adv. Nucl. Phys.}, \textbf{18}:1, 1987.

\bibitem{Buc1}
B. Buck and S.M. Perez, Phys. Rev. Lett. \textbf{50}, 1975 (1983).

\bibitem{Buc2}
B. Buck, A.C. Merchant, and S.M. Perez, Phys. Rev. C \textbf{63}, 037301 (2001).

\bibitem{Per}
S.M. Perez, W.A. Richter, B.A. Brown, and M. Horoi, Phys. Rev. C \textbf{77}, 064311 (2008).


\end{thebibliography}
\end{document}